# Revealing the Atomic-Scale Structure of the Copper–Sulfuric Acid Interface

Lalith Kumar Bhaskar[a#], Sung-Gyu Kang[a,b#], Oliver R. Waszkiewicz[c], Finn Giuliani[c], Baptiste Gault[a,e], Mary P. Ryan[c], Roger C. Newman[d], Gerhard Dehm[a], Rajaprakash Ramachandramoorthy[a,f]*, Ayman A. El-Zoka[c,e]*

*Corresponding authors: r.ram@mpi-susmat.de; ayman.el-zoka@univ-rouen.fr

[#] Equal contribution

[a] Department of Structure and Nano-/Micromechanics of Materials, Max Planck Institute for Sustainable Materials, Düsseldorf, Germany

[b] Department of Materials Engineering and Convergence Technology, Gyeongsang National University, Jinju, Republic of Korea

[c] Department of Materials, Royal School of Mines, Imperial College London, UK

[d] Department of Chemical Engineering and Applied Chemistry, University of Toronto, Canada

[e] Univ Rouen Normandie, CNRS, INSA Rouen Normandie, Groupe de Physique des Matériaux, Rouen, France

[f] Department of Materials Engineering, Faculty of Engineering Technology, KU Leuven, Gebroeders De Smetstraat, Gent, Belgium

**Abstract**

Corrosion originates from atomistic reactions occurring at dynamic solid–liquid interfaces; however, direct experimental observation of these reactions has remained elusive due to the inability to preserve transient interfacial states during characterization. To refine corrosion models, advanced techniques capable of analyzing corrosion interfaces at the atomic scale are essential. Recent advancements in cryogenic-atom probe tomography (cryo-APT) enabled 3D nanoscale analysis of frozen liquid-metal interfaces. However, challenges remain in sample preparation for cryo-APT on metals undergoing corrosion. This study introduces a microcorrosion cell fabricated using localized electrodeposition in liquid (LEL), enabling atomic-scale capture of liquid–metal reactions by integrating picoliter-scale electrolytes encapsulated within sealed metallic microvessels, subsequently analyzed using cryo-APT. This approach enables 3D, nanoscale mapping of corrosion reactions with simultaneous spatial, chemical, and temporal resolution. As a model system, copper exposed to aerated dilute sulphuric acid reveals temperature- and time-dependent interfacial evolution, including nanoscale clustering of copper–sulphate species, enhanced ion pairing at elevated temperature, and the emergence of transient carbon-based interfacial complexes inaccessible to conventional characterization methods. Beyond copper corrosion, the presented microcorrosion cell architecture establishes a strategy for interrogating confined electrochemical and degradation processes across a wide range of material–liquid systems, using a combination of microfabrication and cryo-APT.

## 1. Introduction

Materials degradation and, more specifically, corrosion of metals have proven to continuously impact our economy. The monetary impact of corrosion has been estimated by reports to amount to 2.5 trillion USD per year as of 2013, which is almost 3.4% of the world's annual gross domestic product[1]. Corrosion has an impact on the lifetime of engineering components in different industrial applications[2]. Furthermore, the global goals of achieving sustainable manufacturing and use of materials are also impeded by these corrosion phenomena, as replacing corroded parts accounts for approximately 1.6% to 3.4% of total greenhouse gas emissions[3]. Hence, attempting to design better materials with enhanced corrosion resistance is a necessity.

Copper is one of the most widely used and strategically important metals in modern society due to its outstanding electrical and thermal conductivity[4], mechanical workability[5], and recyclability[6]. These properties make copper essential to the electrification of society, where efficient charge and heat transport are critical for performance and reliability. The global transition toward electrified technologies driven by the rapid expansion of renewable energy systems, electric vehicles, and power infrastructure has led to a sharp increase in copper demand, which is projected to reach approximately 53 million metric tons by 2050[7]. For example, a single electric vehicle contains approximately 80 kg of copper[8], corresponding to several kilometers of wiring, while each megawatt of installed wind-power capacity requires roughly 2.5–3 tonnes of copper for generators and an additional 5.7 tonnes of cables[9]. Beyond energy applications, copper and its alloys are extensively used in heat exchangers, piping networks, electrical wiring, and chemical reactors, where efficient heat and charge transport are critical for system performance. In many industrial environments, copper components are exposed to acidic media during service, which can significantly reduce their corrosion resistance, leading to material degradation. Exposure of copper to acidic media is also very relevant to the microelectronics industry, where electrodeposition of copper interconnects and microscale features are used in the manufacturing of integrated circuits and microchips[10,11]. Early studies by Bockris *et. al.* have shown the importance of acid-copper interactions during the dissolution and electrodeposition process of copper[12,13]. Consequently, understanding the behaviour of copper in acidic environments is essential for improving durability and developing effective corrosion-protection strategies.

Acidic environments, particularly those involving dilute sulphuric acid ($H_2SO_4$), are of particular concern due to the highly corrosive nature of hydrogen ions ($H^+$) and the role of sulphate ions ($SO_4^{2-}$) in modifying copper surface properties[14,15]. In very dilute acid solutions (a few mM), copper corrosion often produces a thin layer of cuprous oxide ($Cu_2O$). This film provides only limited protection against further deterioration, especially with local pH changes or when the solution is mildly acidic [16]. The instability of this oxide layer in dilute sulphuric acid, combined with the continuous dissolution of copper ions, remains poorly understood under realistic conditions[17,18].

Characterizing the role of sulphate ions is also very important, as sulphate ions are perceived to adsorb at the copper/oxide interface, promoting breakdown of the passive cuprite film, and stabilizing localized dissolution by favouring the formation of porous, sulphate-containing corrosion products[19,20]. Moreover, trace impurities and microstructural heterogeneities can further perturb near-surface electrochemical behaviour and oxide film stability, contributing to conflicting observations across different studies[21–23] and underscoring the need for atomic/nanoscale, time-resolved characterization of copper corrosion processes.

Despite extensive research, important gaps remain in understanding the atomic-scale processes at copper surfaces in dilute sulphuric acid, particularly within the near-surface interfacial region, where corrosion is governed by local copper concentrations[24], sulphate adsorption, pH gradients, and transient surface oxide species such as $Cu_2O$ and CuO. Relying on ex-situ methods typically fails to capture the dynamic evolution of electrochemical processes at the copper–electrolyte interface[23].

For decades, metallurgists and corrosion scientists have largely focused on materials degradation by observing changes on the metallic side of the reaction, with less direct visualization of the transformations occurring in the liquid phase. Analysis of liquids in corrosion processes has been largely limited to the use of techniques like mass spectrometry[25] and quartz microbalance systems (QCM)[26]. While QCM is highly sensitive to mass and viscoelastic changes at the solid surface, the interface chemistry is inferred indirectly from the surface-coupled responses. As a result, these approaches lack the spatial and temporal resolution required to resolve atomic- and nanoscale features of the metal–liquid interface where corrosion processes initiate and evolve.

A characterization technique that uses spatially resolved nanoscale mass spectrometry is atom probe tomography (APT). APT has been primarily used in studies of metal and metal oxide systems[27,28]. However, since the 1980s, works by J.A. Panitz and A. Stintz have investigated the possibility of extending the use of APT towards frozen water and aqueous solutions[29–31]. More recent studies on proteins[32] and hydrated glass[33] showed the possibility of developing APT, now known as cryogenic atom probe tomography (cryo-APT), to probe frozen multiphase materials and interfaces. Following this, research has been carried out using cryo-APT to characterize pure water[34], and honey on tungsten needles[35,36]. Systematic, routine investigations of cryo-APT have enabled frozen-atomic-scale analysis of liquid-metal interfaces, with various solute ions detected in frozen solutions in the system containing aqueous NaCl–nanoporous gold[37]. As promising as the use of cryo-APT is for analysing chemically active interfaces, the protocols involved are challenging, primarily due to difficulties for efficiently freezing and preserving solid–liquid interfaces, which are necessary to “lock in” interfacial reactions in real time. As a result, the potential of cryo-APT for quasi-*in situ* studies of corrosion and other chemical reactions has yet to be fully realized.

In this study, we introduce a novel, custom-designed sample platform to enable systematic investigations of copper corrosion in dilute sulphuric acid, examining the effects of both time and temperature. For the first time, we use 3D-printed microcorrosion cells (MCC) fabricated using the localized electrodeposition in liquid (LEL)[38,39] technique tailored for APT analysis, establishing a versatile platform for exploring a wide range of chemical-mechanical reactions on various metals. To bridge existing knowledge gaps, our work delivers near-atomic scale insights into the corrosion of copper in dilute $H_2SO_4$. We monitor the real-time evolution of copper surfaces during corrosion, capturing the dynamic interfacial processes, as well as the nanoscale-level interactions between copper atoms, sulphate ions, and the electrolyte. By freezing these transformations in real time, we reveal the underlying mechanisms of corrosion, the influence of sulphate ions, and the stability of protective oxide films, if any, in dilute acidic conditions. The outcomes of this research will be critical for refining interfacial reaction mechanisms and informing future thermodynamic descriptions of copper–liquid systems. Moreover, the protocols and insights developed here not only deepen our understanding of copper corrosion but also broaden the potential of cryo-APT to investigate other corrosion systems and electrochemical processes of interest.

## 2. Results & Discussion

### 2.1. Microcorrosion Cell Design

To encapsulate dilute sulphuric acid (0.1 M $H_2SO_4$, pH ~0.95) within copper metal, we employed a 3D microfabrication approach based on localized electrodeposition in liquid (LEL)[38,40], which enables the precise printing of micron-scale metallic architectures. A schematic of the fabrication process for the microcorrosion cells (hereafter referred to as MCCs) is presented in Figure 1(a), with additional methodological details available in prior studies[39,41]. Our focus here lies in the optimized design of the MCCs, featuring liquid encapsulation of 0.1 M $H_2SO_4$ within a copper shell, as illustrated in Figure 1(b). The dimensions of these 3D-printed MCCs are 60 microns in diameter and 100 microns in height and the detailed dimensions are provided in Supplementary Figure S1. This configuration facilitates cryo-APT analysis of the liquid–copper interface as a function of time and temperature.

The design requirements for MCCs are primarily governed by the constraints of subsequent cryogenic focused ion beam (cryo-FIB) milling, which remains a challenging stage in cryo-APT analysis of solid–liquid interfaces. The FIB workflows reported so far for frozen liquids require numerous sequential steps, such as freeze-etching the frozen liquid on top of a post, followed by cryo-transfer to FIB for annular milling with progressive current reduction involving long milling times, and a high probability of specimen loss during handling and transfer[33,34,42]. Each step requires meticulous control over the pressure to prevent sublimation or damage. Preparation times averaged between 3–5 hours per specimen, with reproducibility strongly influenced by droplet size, liquid viscosity, and substrate geometry[34].

By precisely defining the geometry and location of the liquid/metal interface within the encapsulated structure, the design eliminates the need for extensive lift-out and site specific positioning. These MCCs are directly printed on a copper substrate of dimensions 5mm × 2mm × 1 mm, as shown in Figure 1(b). This copper substrate is thereafter mounted on an APT sample holder. This design enables direct fabrication of cryo-APT specimens through a single annular milling sequence. As a result, preparation time is reduced to approximately 1–2 hours per tip, and the success rate exceeds 90%.

Ice crystal formation due to atmospheric moisture condensation (frosting), a common artefact reported in earlier cryogenic preparation workflows[34], is effectively suppressed as the solid–liquid interfaces of interest here remain fully encapsulated within the copper shell. Additionally, precise control of the encapsulated liquid volume on the picoliter scale ensures a well-defined metal–liquid interface and reduces the thermal diffusion length within the liquid, enabling more efficient heat extraction through the surrounding copper during rapid freezing.

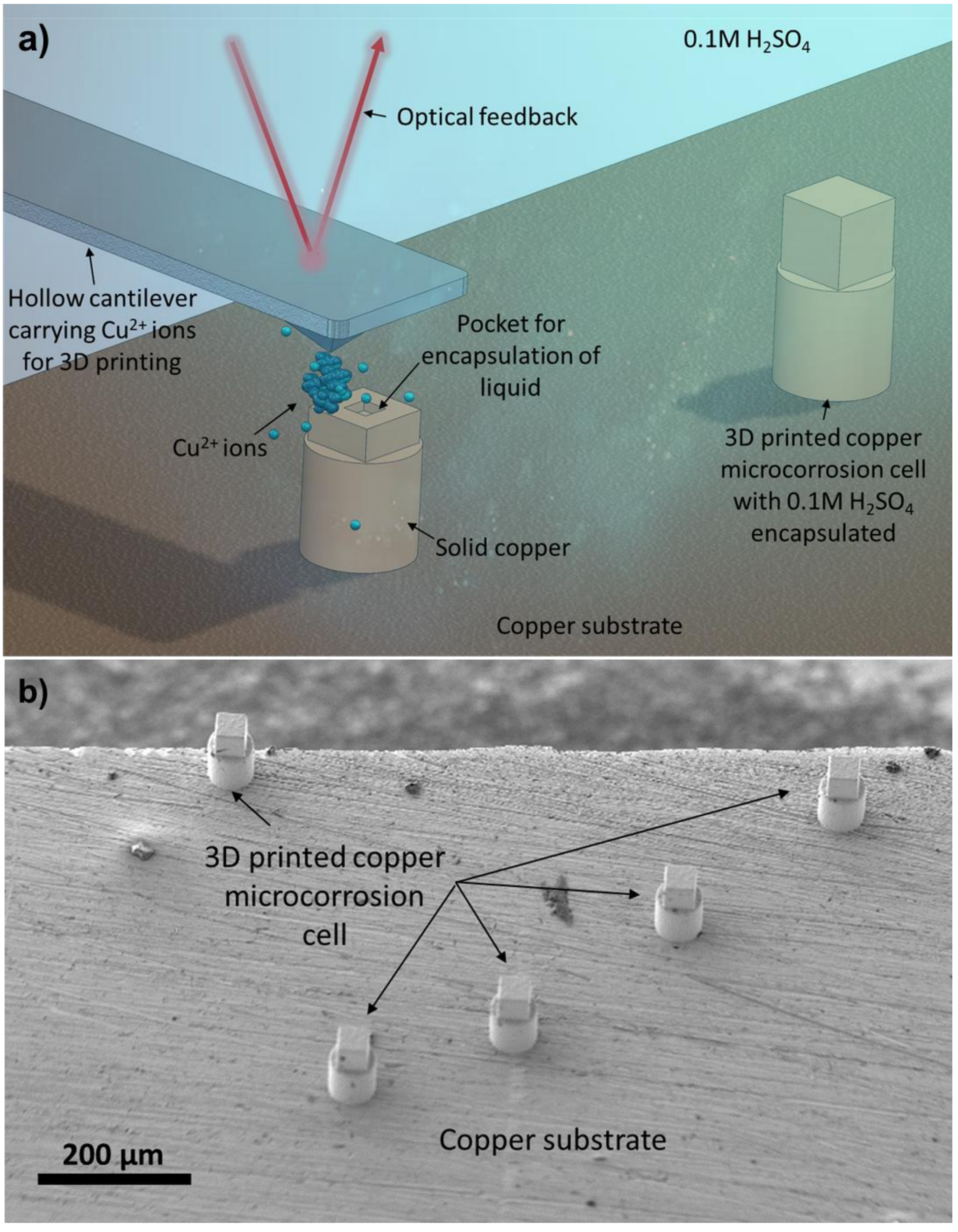

**Figure 1: a) Schematic of the LEL-based microfabrication method for printing MCCs with encapsulated liquid, and b) scanning electron microscopy (SEM) images of the MCCs encapsulating 0.1M $H_2SO_4$ liquid**

As reported in the literature from vitrification studies, increased cooling rates significantly enhance the chances of forming a vitreous solid by suppressing ice nucleation and crystal growth[43]. We also find an inverse correlation between cooling rate and sample volume. For small volumes of liquids, cooling rates exceeding $10^4$ K $min^{-1}$ can be achieved, thereby minimizing crystallization and promoting uniform solidification. Heat transfer simulations (Figure S2) using COMSOL confirm that the MCC design used in this study, enables us to obtain a cooling rate higher than $10^5$ K $min^{-1}$ in the enclosed liquid, even when assuming low heat transfer coefficients of 500 $Wm^2/K$. In the current context, this means that reducing the volume of encapsulated 0.1 M $H_2SO_4$ within the copper cavity improves the uniformity of the frozen interface and limits stress-induced cracking during cryogenic quenching, which will be detrimental during cryo-FIB processing and transfer of the specimen for APT analysis. However, a balance is required; if the liquid volume is too small, the frozen phase may become optically transparent under the electron beam, making it difficult to locate during cryo-FIB milling. Conversely, excessively large volumes reduce the cooling efficiency and can lead to heterogeneous freezing or partial crystallization. Based on these design considerations, an optimized geometry for MCCs ensures both suitable interface positioning for atom probe analysis and maximizing cooling rate during rapid freezing.

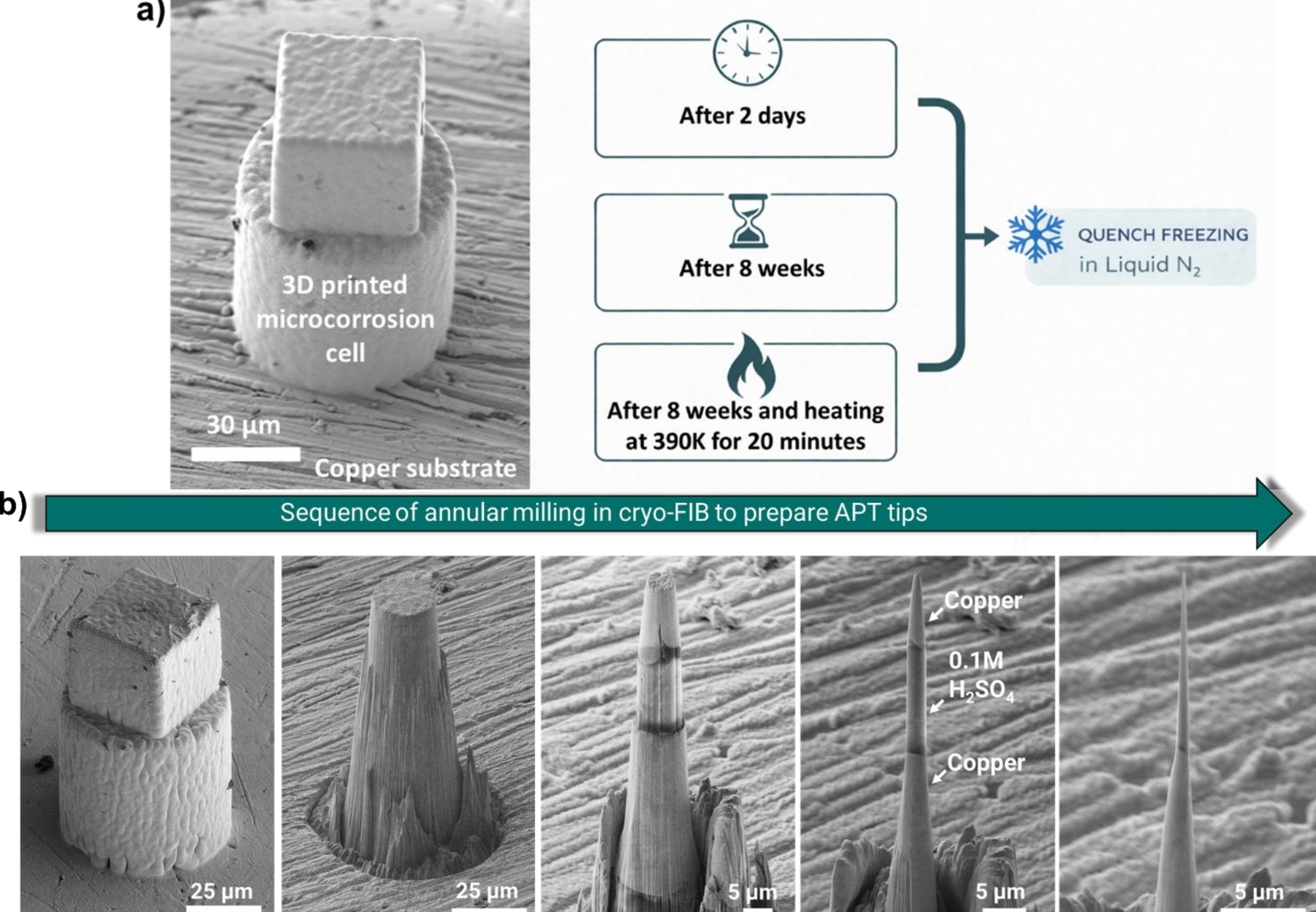

**Figure 2: a) Shows the three different conditions to which the MCC were subjected to before quench freezing; b) Sequences involved in annular FIB milling of a 3D printed MCC into an APT needle**

The successfully microfabricated MCCs were thereafter subjected to three distinct corrosion conditions (as shown in Figure 2(a)): (i) short-term (2 days after printing the MCCs) exposure of copper to sulphuric acid (0.1 M $H_2SO_4$), (ii) long-term exposure of copper to sulphuric acid (8 weeks after printing the MCCs), and (iii) long term exposure with a final heating to 390 K for 20 minutes, followed by rapid quenching in liquid nitrogen in each case. Subsequent APT specimen fabrication was carried out by annular milling of the MCCs using cryo-FIB, using the procedure outlined in the literature[34,37]. Figure 2(b) shows the sequential annular milling process of the MCCs, demonstrating that the liquid (0.1 M $H_2SO_4$)–metal (copper) interface remains intact throughout fabrication.

## 2.2. Chemical Impurities & Inclusions in Electrodeposited Cu

In order to verify the chemical composition and purity of the as-fabricated MCCs copper, the copper layer far from the liquid (0.1 M $H_2SO_4$) – metal (copper) interface was analyzed. Figure 3(a) shows the region of the MCCs from where the APT tip was extracted for further analysis, and Figure 3(b) shows the corresponding atom map of the tip. Figure 3(c) shows the 5 nm horizontal cross-section along with the ion density map and Figure 3(d) shows the through-thickness cross-section extracted from the atom map in Figure 3(b), respectively. Chemical analysis confirms that the 3D microfabricated copper using localized electrodeposition in liquid (LEL) is approximately 99.5% pure copper ions. Trace amounts of $CuO^+$ and $H_2O^+$ were detected throughout the sample at concentrations well below 0.5 ion % in total. Localized relative increase in the presence of $CuO^+$-$H_2O^+$ is marked in the atom maps by a green isosurface. Previous APT investigation of the electrodeposited copper coatings in nuclear waste canisters has identified the oxygen segregation during electrodeposition and the oxidation of grain boundaries[44]. Our cryo-APT method here, confirms the role of entrapped water molecules in eliciting this oxidation.

Ion density maps as shown in Figure 3(c) reveal crystallographic poles and therefore crystal structure of the analyzed copper. The poles observed differ with respect to location above or below the $CuO^+$-$H_2O^+$ rich interface (marked in green), alluding to the likelihood of that interface being a grain boundary. Crystallographic analysis of the various cross-sections of the APT tip in Supplementary Figure S3 confirms this further. We hypothesize that during localized electrodeposition in liquid (LEL), the water molecules are trapped between the printed copper, which could eventually lead to the oxidation of the adjacent copper atoms along grain boundaries, forming nanoscale copper oxide. The observation of entrapped water and the oxidation it could induce highlights an often-overlooked property of electrodeposited metals that merits focused investigation, particularly regarding how electrodeposition parameters influence the amount of trapped water and the extent of grain-boundary oxidation. It should be finally noted here that there were no observations of chloride-related species that might have leaked into the microcell during the electrodeposition (considering that the electrodeposition solution contains chloride ions, and the details of which can be found in section 4.1.).

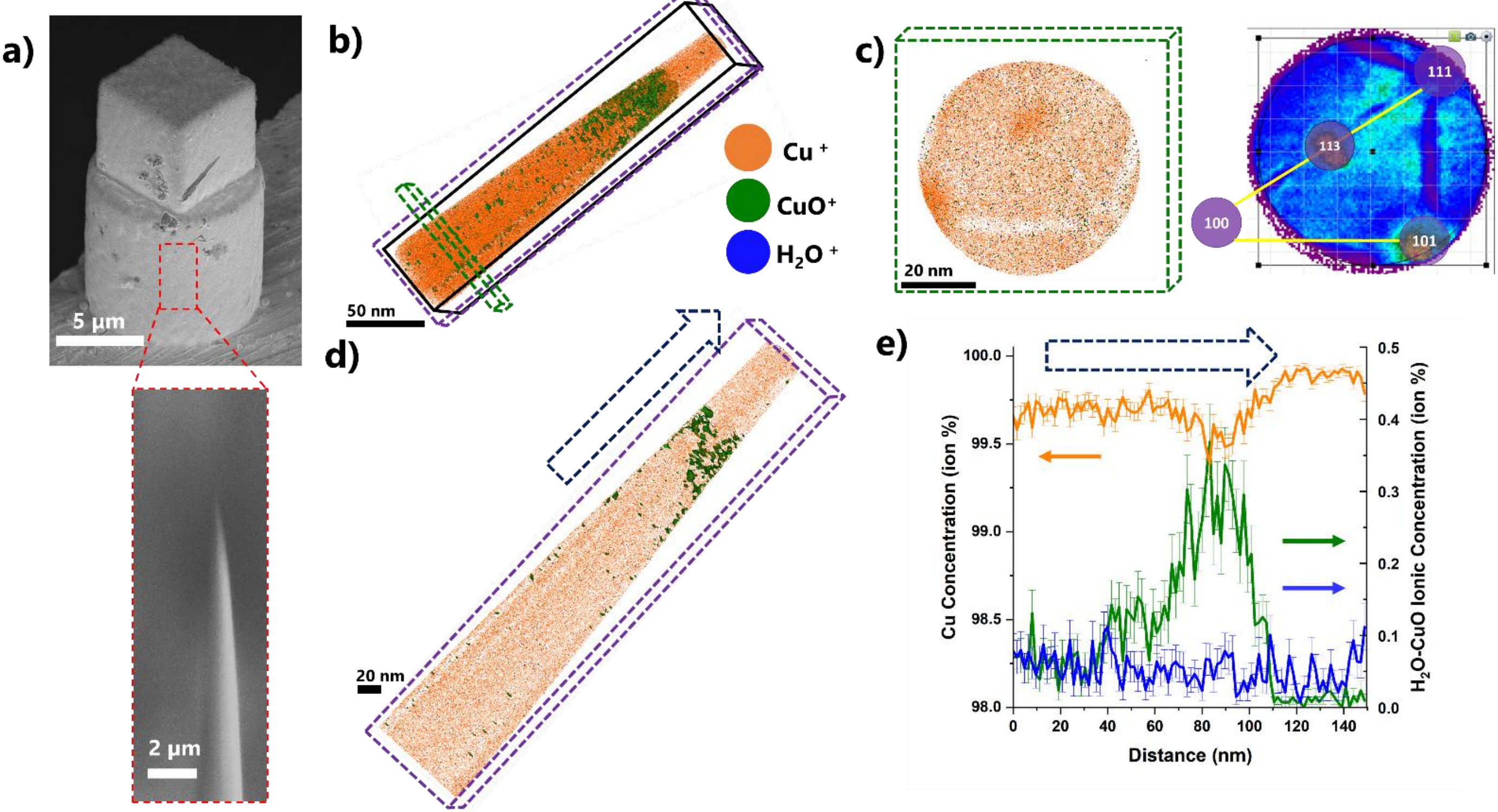

**Figure 3: (a) shows the region of the MCC from which the APT needle was prepared; (b) an atom map of the APT needle, indicating detected copper, copper oxide, and water-related ions; (c) 5 nm cross-section extracted from the atom map in (b), with the corresponding regions marked by color-coded, dashed boundaries also shown is the ion density map revealing crystallographic orientation; (d) a 5 nm through-thickness cross-section of the APT needle shown in (b); and (e) ionic and atomic mapping of the cross-section in (d), highlighting the localization of copper oxide and water molecule along the grain boundary interface**

### 2.3. MCCs 2days after print: Sulphuric acid-copper interface

The development of the liquid (0.1 M $H_2SO_4$) – copper interface is time-dependent, and hence, we investigated the liquid-solid interface in two identical samples, (i) two days and (ii) eight weeks after printing the MCCs. A summary of data acquired for the sample analyzed two days after printing is shown in Figure 4, with Figure 4(a) showing the MCCs from where the APT tip was extracted and the corresponding atom map. The liquid (0.1 M $H_2SO_4$)–copper interface appears more complex, as evidenced by the perpendicular and through-thickness cross-section maps shown in Figure 4(b), as the shape of the interface is not atomically flat, which is also observable from the Supplementary Figure S4. Little pockets (with an apparent opening size of approximately 20 nm, as inferred from atom probe reconstructions using identical parameters across samples) containing high concentrations of $Cu(H_2O)^+$ ions, along with $(H_2O)_n^+$ and $CuSO^+$ species, are observed. Closer chemical mapping of the species across these pockets reveals the presence of $CuSO^+$ species, likely indicating the formation of copper sulphate, with its highest abundance occurring in the central region of the pocket rather than at the corrosion surface, closer to the copper, as seen in Figure 4(b). This suggests that when copper comes into contact with dilute sulphuric acid, layers of copper dissolve into the solution, producing copper sulphate, with dissolved oxygen acting as an initial oxidant for the corrosion reaction. This pronounced clustering observed for copper sulphate in the liquid phase can be attributed to strong electrostatic interactions between $Cu^{2+}$ and $SO_4^{2-}$ ions. The ion activity coefficients of $CuSO_4$ are significantly lower when compared to other copper salts or sulphates across the entire concentration range (e.g., 0.158 at 0.1 M $CuSO_4$), indicating substantial ion pairing[45]. This strong pairing arises from the high charge density of both $Cu^{2+}$ and $SO_4^{2-}$, which leads to enhanced coulombic attraction compared to monovalent systems. In addition, sulphate can coordinate $Cu^{2+}$ through its oxygen atoms, forming stable contact or inner-sphere ion pairs that partially replace water in the $Cu^{2+}$ hydration shell with little energetic penalty. Together, these effects weaken effective electrostatic screening and enhance ion–ion correlations in this 2:2 electrolyte within the liquid phase, leading to observable Cu-$SO_4$ clustering as observed in Figure 4(b).

Also, hydrated copper ions in the pocket suggest copper ions dissolved in the solution and subsequently frozen. This observation is consistent with prior cryo-APT studies made by El-Zoka et al.[37]. In that work, Ag was detected as both atomic ions and Ag-containing molecular ions coordinated with one or more $D_2O$ molecules, i.e., $Ag(D_2O)^+$ and $Ag(D_2O)_2^+$, within the frozen liquid phase. These hydrated ionic species were found not within the metallic Ag but within the frozen water confined on top of the nanoporous gold, strongly suggesting that they originated from Ag ions dissolved in solution before freezing, rather than from field-evaporated metallic silver. Nevertheless, it should be noted that our observations of the Cu-S-O complex ions could also form as a result of field evaporation-induced interaction between Cu ions, and sulphate ions in the solution. Other copper-sulphur species were observed within the liquid phase, including $CuS_xH_yO^+$, as shown in Supplementary information Figure S5; however, their abundance was less than 5 ion % within the liquid phase as observed from the ionic percentage map in Figure 4(c). It should be noted that certain peaks in the APT spectrum index to both $CuO^+$-based compositions alongside $CuS_x^+$ species; however, the AP Suite analysis software indicates a higher probability for sulphur-based compounds. Please refer to Supplementary

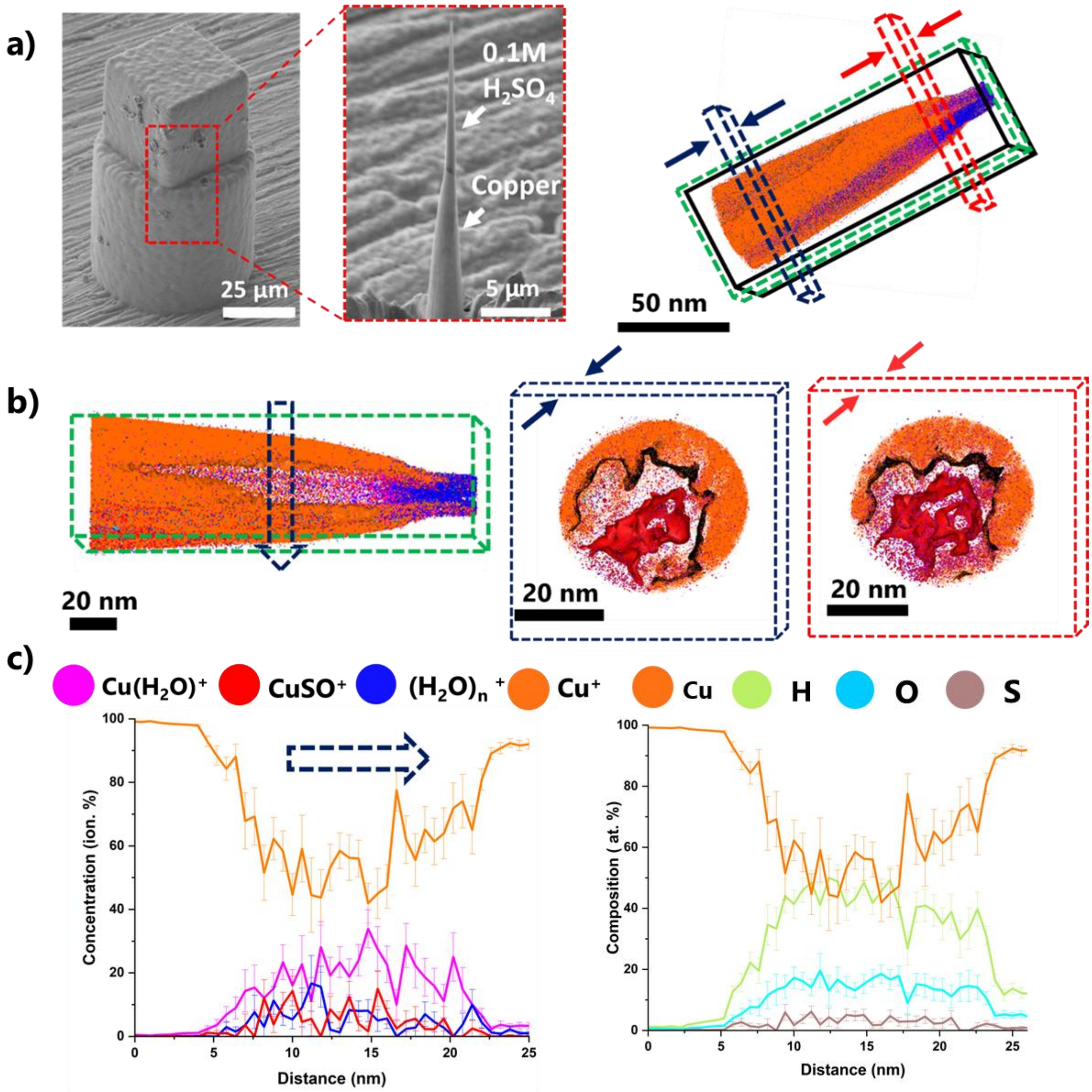

**Figure 4: (a) shows the region of the MCC (after two days) from which the APT needle was prepared and also shown is the atom map of the corresponding APT needle at frozen liquid (0.1 M $H_2SO_4$)–copper interface; (b) a 5 nm perpendicular and through-thickness cross-sections extracted from the atom map in (a), with the corresponding regions marked by color-coded, dashed boundaries. The cross-section reveals the various complex copper sulphur species as highlighted in the legend; and (c) ionic and atomic mapping of the through-thickness cross-section in (b), highlighting the concentration of $Cu(H_2O)^+$ and $CuSO^+$ ions**

Information, Figure S5.

## 2.4. MCCs 8 weeks after print: Sulphuric acid-copper interface

A second set of MCCs containing picoliters of 0.1 M $H_2SO_4$ was left under atmospheric conditions for eight weeks after microfabrication to investigate the temporal evolution of the liquid–metal interface composition. Cryo-APT analyses of MCCs examined within two days of fabrication showed no evidence of copper oxide complexes at the interface. Given that previous studies have reported substantial growth of copper oxide films by several hundred nanometers after days to weeks of ambient exposure[46], the prolonged holding period in this study was to determine whether such oxide growth could be detected. A summary of the corresponding data analysis is presented in Figure 5. Figure 5(a) shows the atom map of the APT tip, while Figure 5(b) presents a 5 nm-thick cross-section taken perpendicular to the atom map in Figure 5(a). A complementary through-thickness cross-section parallel to the APT needle axis is provided in Supplementary Figure S6. In contrast, the penetration depth was limited to 20–30 nm in MCCs analyzed after 2 days. It is important to note that, from prior APT studies of pores and fluid inclusions, a true volumetric distribution cannot be determined with high certainty, as trajectory aberrations during cavity opening can project ions originating at the interface into the apparent interior of such features[47,48]. Therefore, the absolute dimensions of these features are subject to reconstruction artifacts. However, all datasets were reconstructed using identical parameters, ensuring that the observed differences in penetration depth are internally consistent and meaningful for comparative analysis.

As seen in Figure 5(c), hydrated copper ions were observed to increase in concentration in the liquid phase to ~50 ion %, compared to ~30 ion % for MCCs analyzed after 2 days. Similarly, the concentration of copper sulphate species increased from ~10 ion % to ~15 ion % in the liquid phase over the same time period. This increase in both hydrated copper ions and copper sulphate species from 2 days to 8 weeks suggests that corrosion in the presence of entrapped liquid follows an exponential decay-like behaviour, with a high initial corrosion rate that gradually decreases over time.

The formation of copper sulphate has previously been observed only through energy-dispersive X-ray spectroscopy (EDS) in the SEM[49], where higher concentrations of sulphuric acid were necessary to increase the likelihood of detection of copper sulphate formation. In this case, even with picoliters of dilute 0.1 M $H_2SO_4$, it was possible to detect the copper-based compounds. Additionally, it is worth noting that EDS techniques have inherent limitations in accurately detecting light elements, such as sulphur and oxygen[50].

One interesting aspect is that even in the MCCs analysed after a couple of weeks, no evidence of copper oxide formation was detected at the liquid–metal interface. The hypothesis to the observation is that the encapsulated liquid phase consists of dilute sulphuric acid (0.1 M $H_2SO_4$) printed under aerated, but not oxygen-saturated, conditions. Under such mildly oxidizing yet acidic environments, copper dissolution is thermodynamically favoured over oxide formation, which is also supported by the Pourbaix diagram (also shown in Supplementary Figure S9). From literature[51], the presence of dissolved oxygen in sulphuric acid enhances the oxidation of $Cu^{+}$ to soluble $Cu^{2+}$ species, promoting the formation of $CuSO_4$ rather than the stabilization of copper oxide ($Cu_2O$) films.

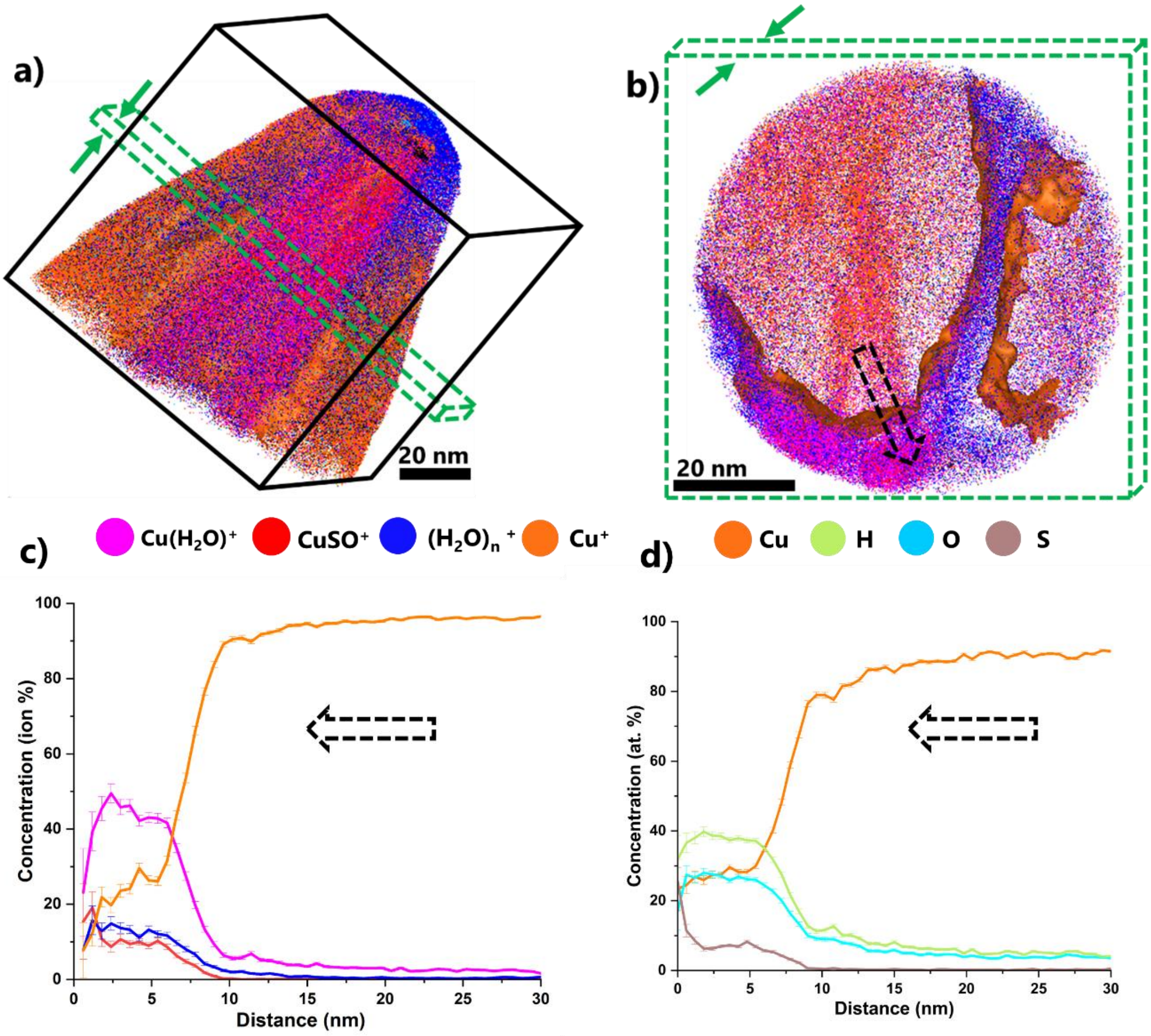

**Figure 5: a) Atomic map extracted within the frozen liquid (0.1 M $H_2SO_4$) – metal (copper) interface of MCCs microfabricated and cryo-FIB milled after exposure for 8 weeks; (b) a 5 nm cross-section of the map in (a) showing the copper complex, and water ions, with the corresponding regions marked by color-coded, dotted boundaries; c) ionic and d) atomic mapping of the thin cross-section in b) showing the formation of hydrated copper ions and copper sulphate ions**

## 2.5. Role of Temperature: Carbonated copper oxide at the sulfuric acid-copper interface

To examine how elevated temperature influences the interaction between copper and sulphuric acid, MCCs were heated to 390 K for 20 minutes in air, then rapidly frozen in liquid nitrogen before cryo-APT analysis. Above this temperature, stress cracking was observed in similar 3D-printed copper microarchitectures[40]. From the atom maps in Figure 6 (a) and 6(b), the rate of copper corrosion appears to be highest in this case, as ionic and atomic mapping in Figure 6(c) and 6(d) respectively, reveals the lowest abundance of elemental copper in the interface next to liquid-corroded region, with just ~30-40 atomic percentage compared to ~90 atomic percentage in previous cases. Correspondingly, the highest concentration of hydrated copper ion species, with a maximum ion percentage of ~45%, is observed. Presence of copper sulphate, approximately 10 ions %, is observed in the chemical map in Figure 6(c). This observation is consistent with high-temperature spectroscopic studies showing that $Cu–SO_4$ association remains thermodynamically favourable within the liquid phase and becomes stronger with increasing temperature, reflecting enhanced ion pairing rather than dissociation[45]. The corresponding through-thickness cross-section parallel to the reconstructed APT data set is shown in Supplementary information Figure S7.

Also observable from Figure 6(c) is evidence of a carbonated copper oxide compound at the solid–liquid interface. We hypothesize the formation of these $CuOC^+$ (carbonated copper oxide compound) species to the role of dissolved carbon dioxide ($CO_2^+$) in the sulphuric acid solution. $CO_2^+$ peak is detected in the mass spectra of all MCCs samples (Supplementary Figure S8), but at consistently low concentrations (≈0.120–0.150 ion %), indicating that carbon-bearing species are present only in trace amounts. The absence of a peak at m/z = 12 ($C^+$) across all samples further rules out free carbon contamination. We hypothesize that $CO_2^+$ is a result of dissolved carbon dioxide in the printing solution and not a formation that arises solely from heating to 390 K.

Notably, an increase in the detected $CuOC^+$ species is observed only after heating to 390 K, and this species preferentially accumulates at the liquid–metal interface. In addition, the $CuOC^+$ concentration within the liquid region also increases from 0.193 ion % in the as-received MCC to 2.12 ion % in the heated MCC, as also observable from the APT mass spectrum in Fig. 6(e). Finally, in addition to $CuOC^+$, we could also detect $CuCO_2^+$ ion, but only in the sample heated to 390 K, and it is confined to the liquid region.

On continuing with the hypothesis, the appearance of the $CuOC^+$/$CuCO_2^+$ ionic species only after heating the MCC to 390 K is most plausibly attributed to a temperature-enabled interfacial chemistry in the presence of $CO_2^+$ as a source of carbon. Sulphuric acid undergoes a second, incomplete dissociation according to,

$$(HSO_4^- \rightleftharpoons H^+ + SO_4^-) \qquad \text{--(1)}$$

has a measurable acidity constant that is moderately weak at standard conditions. At 298 K, the second dissociation constant ($Ka_2$) corresponds to a $pKa_2$ of about 1.9–2.0[52], meaning only a fraction of $HSO_4^-$ dissociates to release a proton under typical conditions. Importantly, the temperature dependence of this equilibrium is such that $pKa_2$ increases with temperature (i.e., $Ka_2$ decreases) because the dissociation is exothermic; experimental correlations for $HSO_4^-$ show empirically that $pKa_2$ rises as temperature increases. For example, applying a standard temperature-dependent expression for $pKa_2$ yields approximate estimates of $pKa_2 \approx 2.0$ at 298 K and $pKa_2 \approx 2.3–2.5$ at ~390 K[53] (depending on ionic strength and concentration), indicating less second dissociation and thus relatively fewer free protons at high temperature. This means that locally the pH could vary significantly and may no longer be ≈0.95 and could be higher at higher temperatures.

Within this context, inspection of the Cu–$H_2O$ Pourbaix diagrams at both 298K and 390K (Supplementary Figure S9) reveals that the stability boundary for $CuOH^+$ shifts toward lower pH (from pH ~ 7 to ~3) with increasing temperature. Although these diagrams do not explicitly account for the presence of sulphate or carbon-containing species, their omission is expected to further modify interfacial speciation under the present experimental conditions. Consequently, at elevated temperature, local changes in proton activity and electrochemical gradients at the metal–liquid interface may facilitate the transient formation of Cu–O(H) species. These Cu–O surface sites can then preferentially interact with dissolved $CO_2$ derived species present in the encapsulated liquid, leading to the formation of short-lived oxycarbonate-like Cu–O–$C(O)_x$ complexes at the hot interface. During atom probe analysis, such interfacial complexes are plausibly detected as $CuOC^+$ and, in the liquid region, as $CuCO_2^+$ molecular ions.

While this mechanistic interpretation necessitates further targeted experimental/computational validation, the present observations underscore the most important novelty of the work: interfacial reactions between metals and liquids can be successfully captured, arrested, and investigated at atomic resolution using cryogenic APT. Importantly, this approach enables the detection of transient ionic species and interfacial complexes that are not predicted by equilibrium Pourbaix diagrams or conventional thermodynamic models. And such a methodology demonstrated in this work provides a powerful experimental framework for refining interfacial reaction mechanisms and informing future thermodynamic descriptions of metal–liquid systems under non-equilibrium conditions.

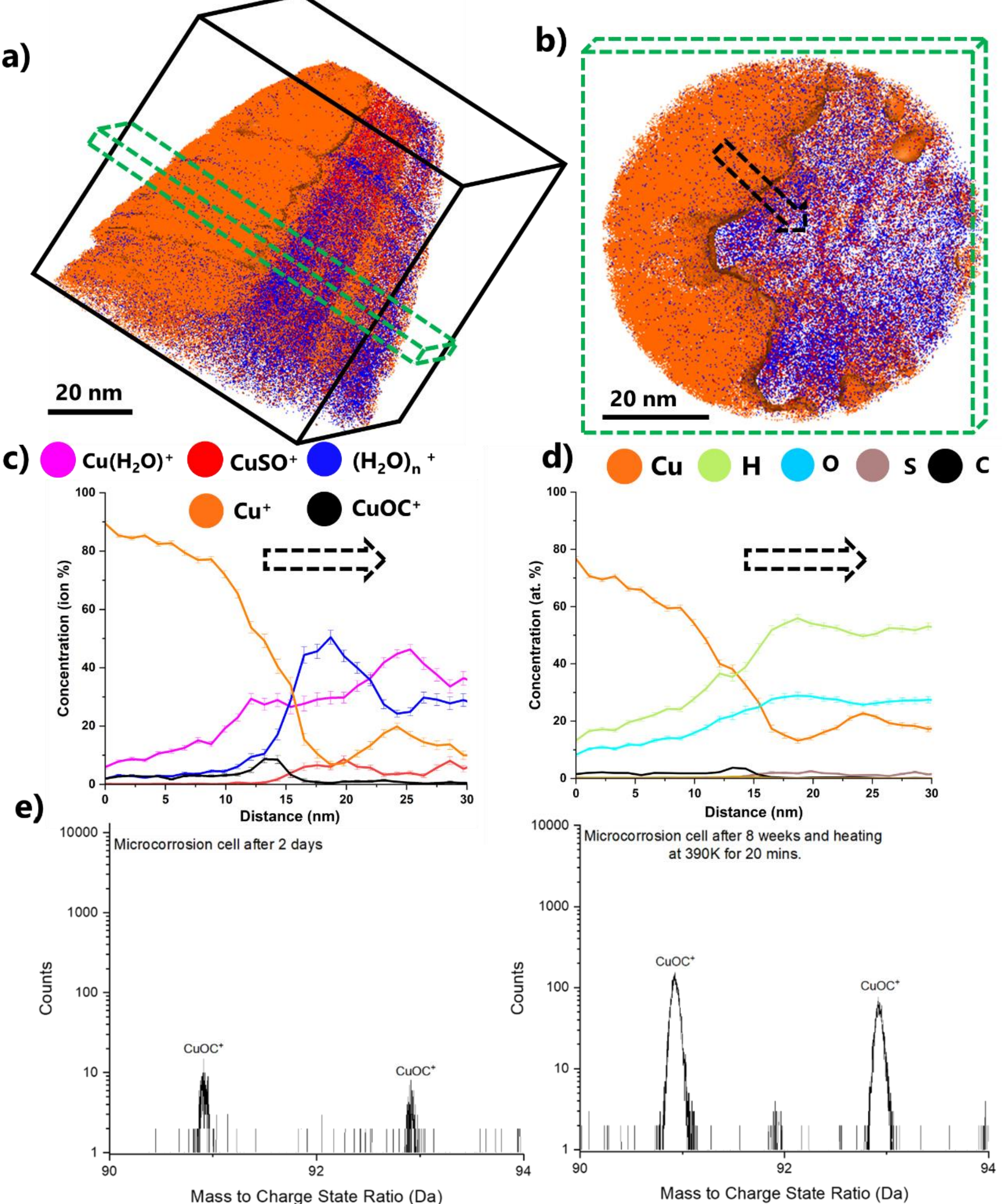

**Figure 6: a) Atomic map extracted within the frozen liquid (0.1 M $H_2SO_4$) – metal (copper) interface of MCCs microfabricated and cryo-FIB milled after exposure for 8 weeks and thereafter heated at 390K for 20 minutes; (b) a 5 nm cross-section of the map in (a) showing various copper complexes and water ions, with the corresponding regions marked by color-coded, dotted boundaries; c) ionic and d) atomic mapping of the thin cross-section in b); and e) mass spectrum of $CuOC^+$ ions obtained from MCCs investigated after two days and the MCCs subjected to heating at 390K for 20 minutes after 8 weeks of printing**

## 3. Conclusions

In this work, a novel 3D-printed microcorrosion cell design that enables efficient, high-yield cryo-APT analysis of solid–liquid interfaces is demonstrated here for the analysis of copper corrosion in aerated, dilute sulphuric acid (0.1M $H_2SO_4$). This approach achieves precise control over liquid volume and interface positioning, reducing cryo-APT sample preparation time while delivering greater than 90% success rates. Atomic scale mapping revealed that corrosion initiated through the formation of copper sulphate, containing water pockets within the first 10–20 nm of the copper surface. These nanoscale pockets acted as localized reaction sites, with their size and distribution decreasing with depth, indicating heterogeneous corrosion across the APT specimens. Time-resolved experiments show that prolonged exposure (8 weeks) increases, though not significantly, both hydrated copper ions and copper sulphate ions at the interface, with liquid inclusions penetrating deeper into the copper phase. This pointed towards a progressive interface roughening and localized attack rather than uniform surface dissolution. Also, in the case of both the microcorrosion cells analyzed, there was no evidence of copper oxide at the liquid-metal interface, in line with what has been reported in the literature. Analysis of elevated-temperature (390 K) MCCs showed a significant increase in corrosion, yielding the highest concentrations of hydrated copper ions, alongside carbonated copper oxide deposits observed only under high-temperature conditions, suggesting temperature-enabled interfacial chemistry in the presence of dissolved CO2 as a carbon source. Beyond providing fundamental insights into the role of sulphate ions, temperature, and dissolved carbon dioxide in copper degradation, this work establishes a versatile platform for in situ, atomic-scale studies of metal–liquid interactions capable of capturing the corrosion process in intermediate metastable states. The methodology can be readily extended to investigate corrosion and electrochemical processes in a wide range of metals and encapsulated solutions, offering a path toward better-informed design of corrosion-resistant materials and protective strategies.

## 4. Methods

### 4.1. Printing of microcorrosion cells

In this work, a localised electrodeposition in liquid (LEL) technique was employed to fabricate microcorrosion cells using a force-controlled electrodeposition system (CERES, Exaddon AG, Switzerland). The electrolyte for printing the structure consisted of 0.5 M $CuSO_4$ in 51 mM $H_2SO_4$ and 0.48 mM HCl with brightener and leveler additives, supplemented with additives to enhance brightness and refine the grain structure of the deposited copper. This solution was held in a chamber and delivered through a 300 nm-diameter orifice in a silicon nitride AFM cantilever tip, which was immersed in a standard three-electrode electrochemical cell containing the test solution (0.1 M $H_2SO_4$ in this study). Electrodeposition was performed under an applied pressure of 50 mbar, on a diamond-polished polycrystalline copper substrate (5 mm × 2 mm × 1 mm) attached to the working electrode using silver paste, mounted on a Si/Ti (10 nm)/Cu (100 nm) support. The small substrate dimensions were selected to facilitate and ease the preparation of atom probe tomography (APT) needles via cryo-focused ion beam (cryo-FIB) milling. Deposition proceeded via a voxel-by-voxel approach. Each time a growing copper voxel caused cantilever deflection (detected via laser deflection signal), the AFM tip advanced to the next voxel coordinate. The 3D geometry was first designed in SolidWorks, then processed through the Exaddon voxeliser software for suitable voxelisation. Further details of the LEL process can be found elsewhere[41].

### 4.2. FIB sample preparation

Atom probe tomography is very useful in obtaining chemical analysis at the near atomic scale[28,54]. However, conventional Ultra-high vacuum-cryogenic sample transfer is required to keep the sample frozen in its native state and to limit any frosting from residual water vapour. This transfer mechanism is detailed in previous cryo-APT studies, pioneered in the Laplace project from the Max-Planck-Institute for Sustainable Materials, in Düsseldorf[55]. A transfer suitcase maintained at ultra-high vacuum ($10^{-11}$ mBar) and cryogenic temperatures (≈ 90 K) from Ferrovac was used to transport the frozen samples between an inert nitrogen glovebox, a Thermo Fisher Hydra/Helios dual beam PFIB (Xe plasma source) fitted with a cryo-stage, and Cameca LEAP 5000 XR (Cameca Instruments). The samples were rapidly frozen inside the glovebox by plunging into liquid nitrogen. The frozen samples were then transferred from the glovebox into the transfer suitcase via a load lock, which reached a vacuum of $10^{-6}$ mBar. For the sample heated at 120 ºC, a convection air furnace was used. The sample was placed into the furnace, 20 mins after reaching 120 ºC, the sample was taken out and quenched in a liquid nitrogen bath immediately. The sample was then transferred into the glovebox within the liquid nitrogen bath, followed by transfer to the FIB. APT tips of liquid-metal interface were created in the Thermo Fisher Hydra PFIB (Plasma Focussed Ion Beam) following the same procedure as reported by El-Zoka *et. al.*[37] and Stender *et. al.*[34]. Ion beam currents ranged from 60 nA to 0.1 nA with a voltage of 30 kV.

### 4.3. Cryo-Atom Probe Tomography (cryo-APT)

All samples were transferred from the FIB to the APT under HV (high vacuum, < $10^{-6}$ Pa) and cryogenic temperature (-196°C) in a Ferrovac suitcase. APT analysis was carried out in a Cameca LEAP 5000 XR (Cameca Instruments) using Laser pulsing mode with a pulse of 60 pJ and voltage mode pulse rate of 50 kHz. The detection rate was varied between 0.1-1% (0.001-0.01 ions per pulse). The stage temperature was kept constant at 50 K for all experiments. Reconstructions were carried out in AP Suite 6.1, IVAS 6.1.3.42.

## 5. Acknowledgements

L.K.B. and R.R. would like to acknowledge funding from the European Research Council (ERC) (Starting grant agreement No. 101078619; AMMicro). Views and opinions expressed are however those of the author(s) only and do not necessarily reflect those of the European Union or the European Research Council. Neither the European Union nor the granting authority can be held responsible for them L.K.B. acknowledges partial funding from the SFB 1394 (project ID 409476157). G.D. and L.K.B. are grateful for support by KSB Stiftung (project no 3.2025.14). S-G.K. acknowledges funding from the National Research Foundation of Korea (NRF, RS-2026-25486157). A.A.Z acknowledges support from the Department of Materials at Imperial College London. Authors are thankful for access to the Imperial Cryogenic Microscopy Centre in Imperial College London, which is supported by EPSRC grant no. EP/V007661/1**.**

## 6. Conflict of Interest

No conflicts of interest to be declared.

# Supplementary information

## Revealing the Atomic-Scale Structure of the Copper–Sulfuric Acid Interface

Lalith Kumar Bhaskar[a#], Sung-Gyu Kang[a,b#], Oliver R. Waszkiewicz[c], Finn Giuliani[c], Baptiste Gault[a,e], Mary P. Ryan[c], Roger C. Newman[d], Gerhard Dehm[a], Rajaprakash Ramachandramoorthy[a,f]*, Ayman A. El-Zoka[c,e]*

*Corresponding authors: r.ram@mpi-susmat.de; ayman.el-zoka@univ-rouen.fr

[#] Equal contribution

[a] Department of Structure and Nano-/Micromechanics of Materials, Max Planck Institute for Sustainable Materials, Düsseldorf, Germany

[b] Department of Materials Engineering and Convergence Technology, Gyeongsang National University, Jinju, Republic of Korea

[c] Department of Materials, Royal School of Mines, Imperial College London, UK

[d] Department of Chemical Engineering and Applied Chemistry, University of Toronto, Canada

[e]Univ Rouen Normandie, CNRS, INSA Rouen Normandie, Groupe de Physique des Matériaux, Rouen, France

[f] Department of Materials Engineering, Faculty of Engineering Technology, KU Leuven, Gebroeders De Smetstraat, Gent, Belgium

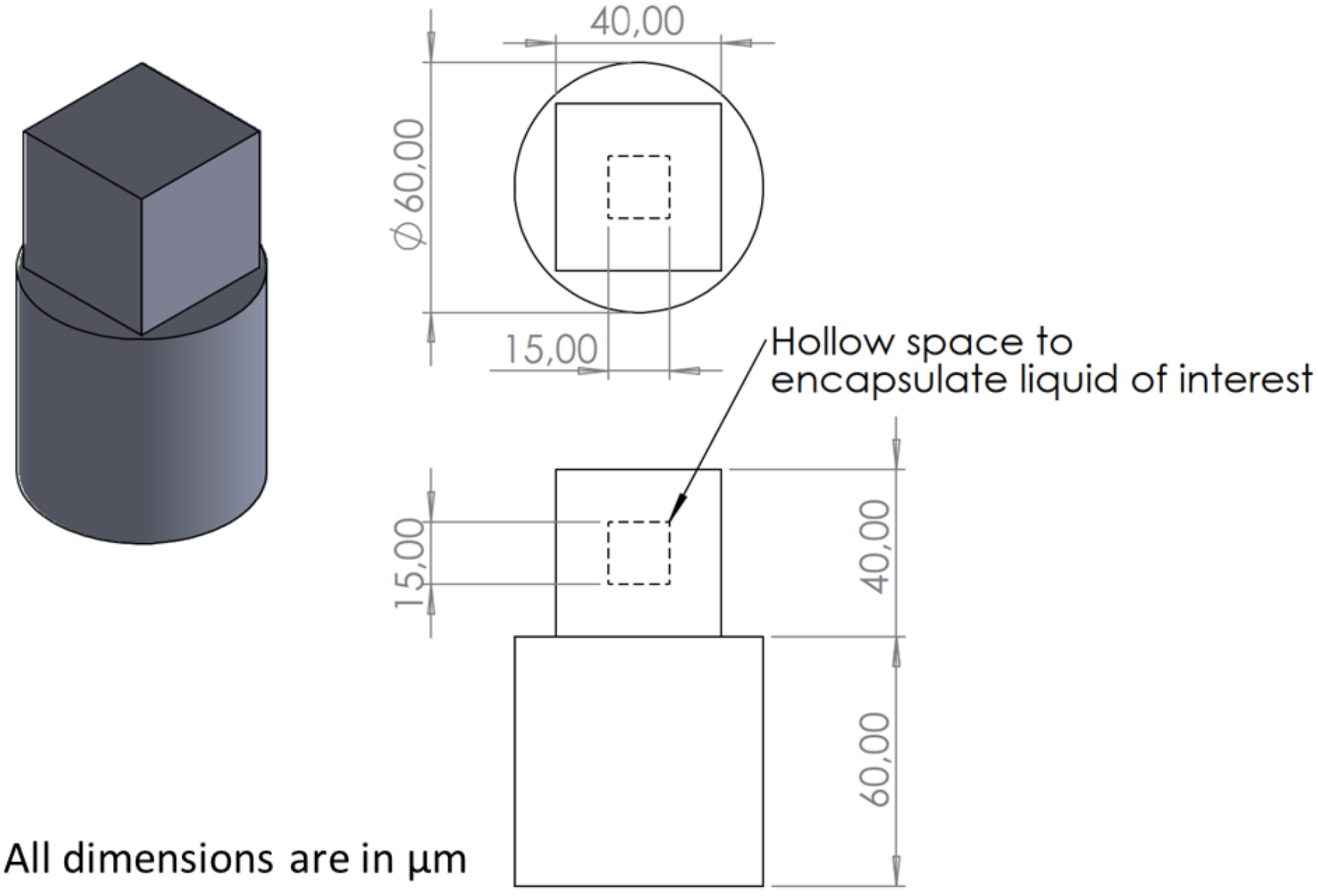

**Figure S1** shows the CAD model of the optimised geometry for the micro-corrosion cell

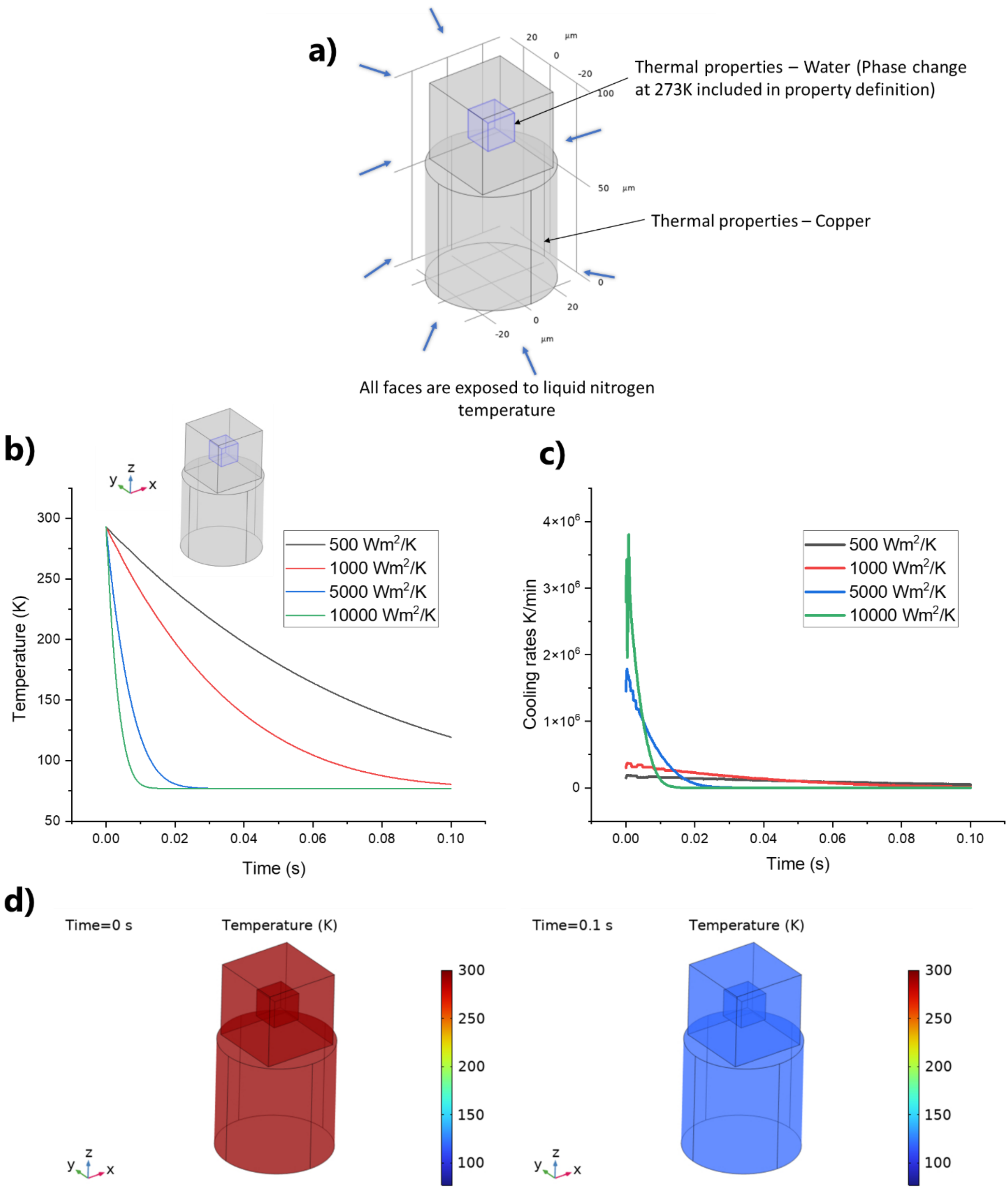

**Figure S2. (a)** Boundary conditions applied to the microcorrosion cell in the heat transfer simulations performed using COMSOL. **(b)** Simulations were conducted for a fixed duration of 0.1 s while varying the heat transfer coefficient at the solid–liquid interface; the temperature evolution over 0.1 s was extracted from the region highlighted in the inset. **(c)** Corresponding cooling rates estimated for different interfacial heat transfer coefficients. **(d)** Representative temperature distribution within the microcorrosion cell at 0 s and 0.1 s for a heat transfer coefficient of 500 W m$^{-2}$ K$^{-1}$.

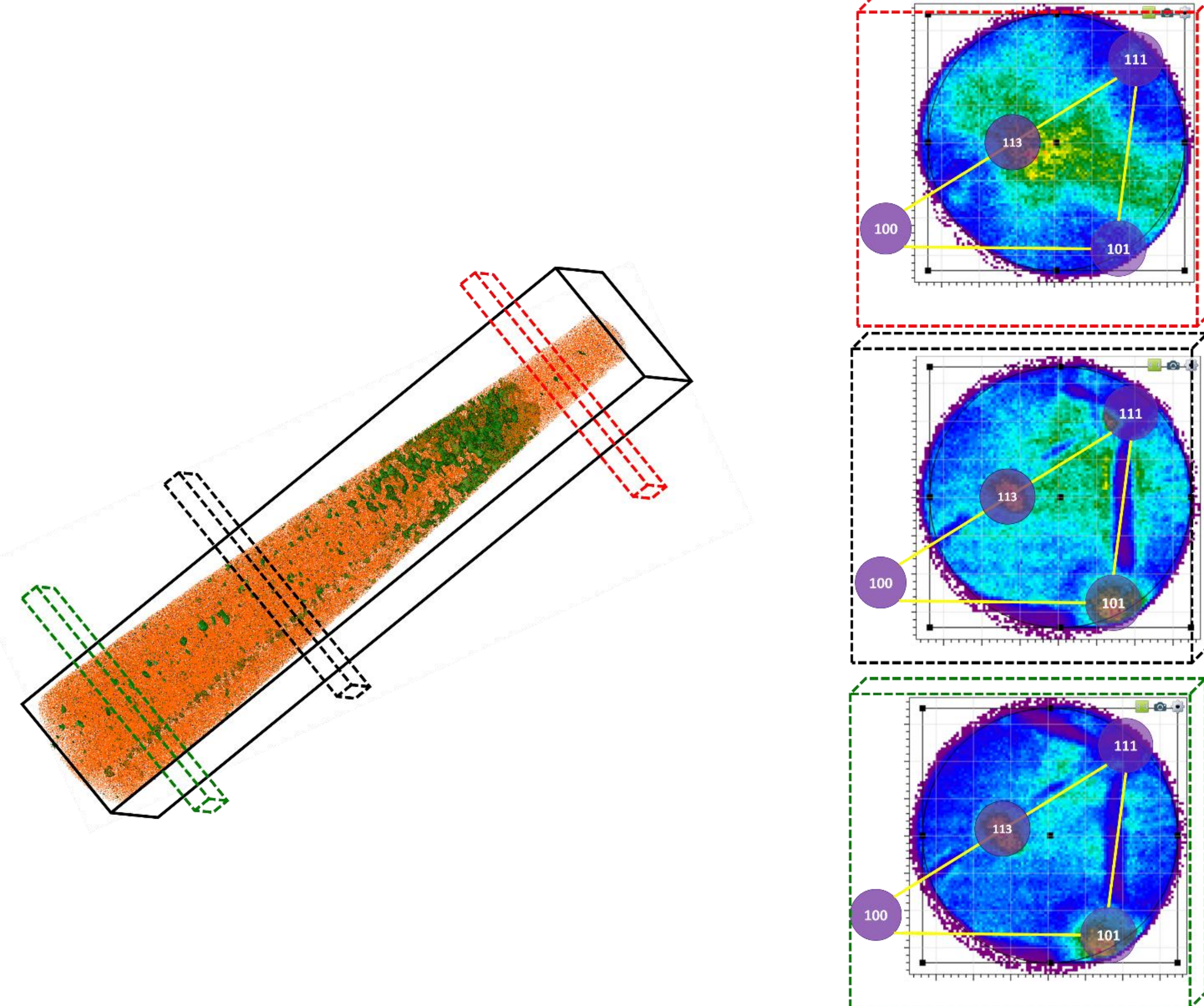

**Figure S3** Ion density maps, along different cross-section of the APT tip as shown in main Figure 3(b). Crystallographic orientation indexing of the various cross-sections shows differences in orientation between the regions below and above the grain boundary.

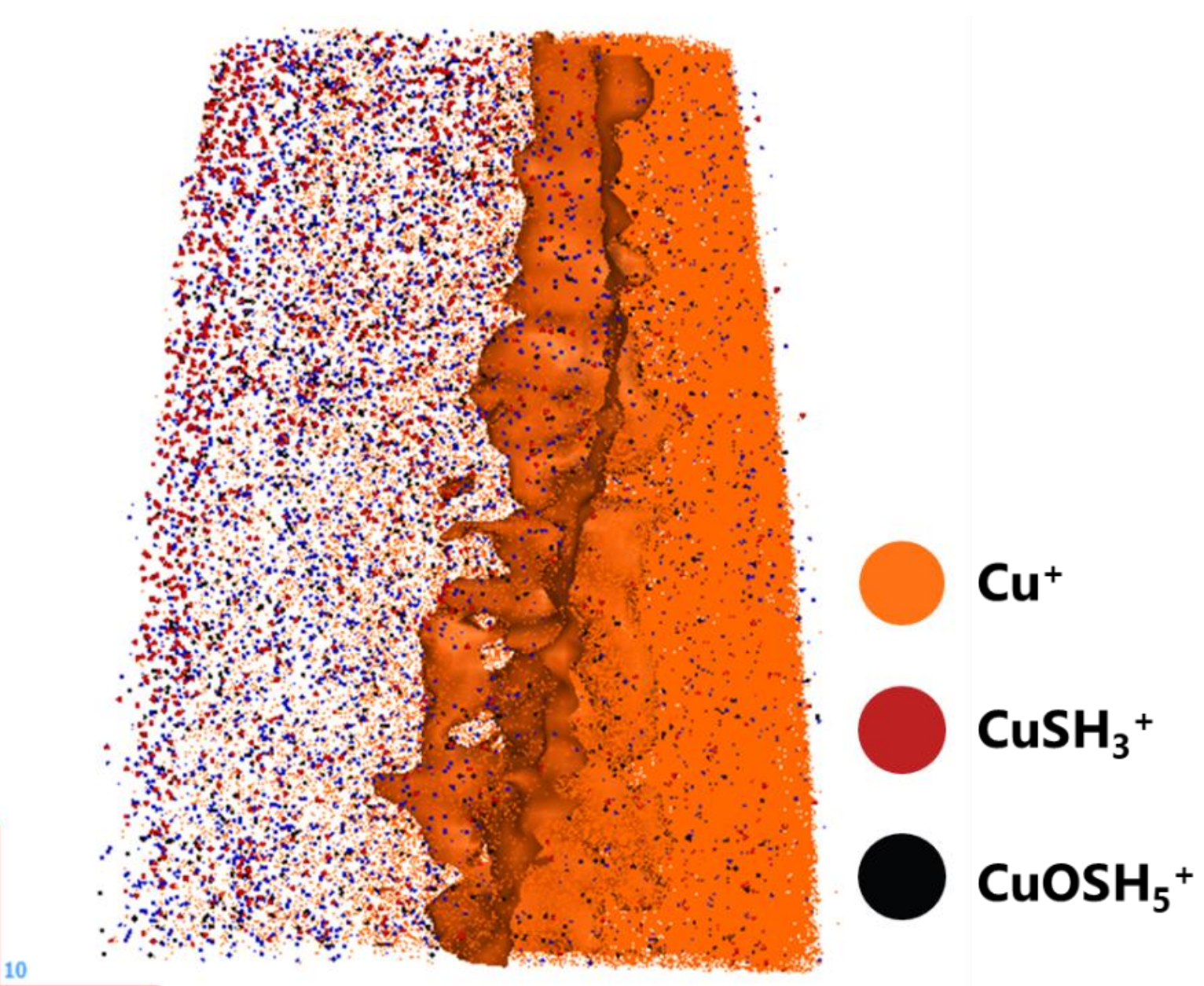

**Figure S4** Through-thickness cross-section parallel to the APT needle for the microcorrosion cell fabricated and cryo-FIB milled after exposure for 2 days

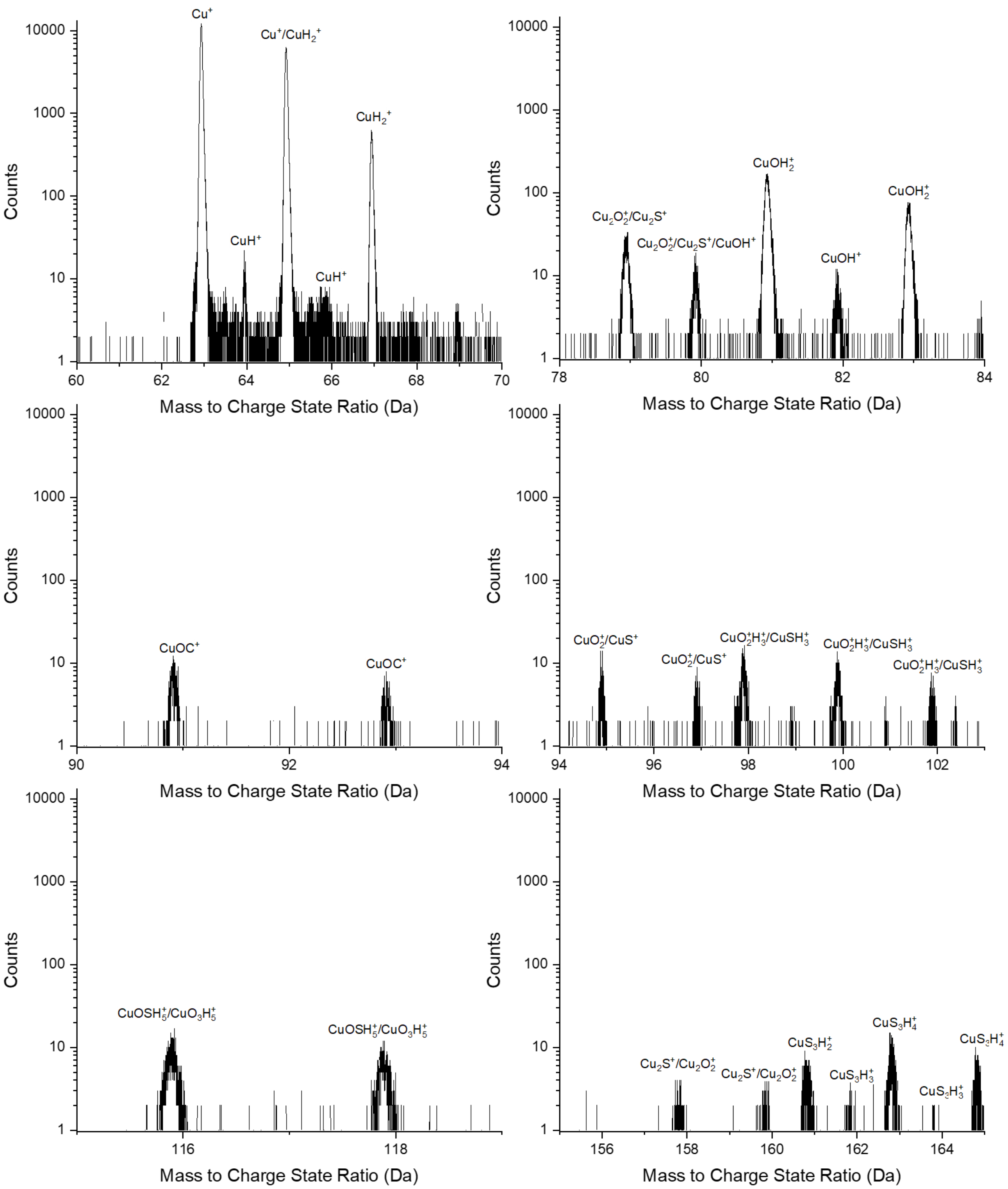

**Figure S5** shows Mass spectra for all major compositions identified in the liquid phase. In addition to the dominant copper–sulfur species, minor copper–sulfur–oxygen/hydrogen species ($CuS_xH_yO$) were also detected, as shown in Supplementary Information Fig. S2, though their abundance within the liquid phase was below 5%. Certain peaks in the APT spectrum can be assigned to both $CuO^+$-based and $CuS_x$ species; however, AP Suite analysis indicates a higher probability for sulfur-based assignments.

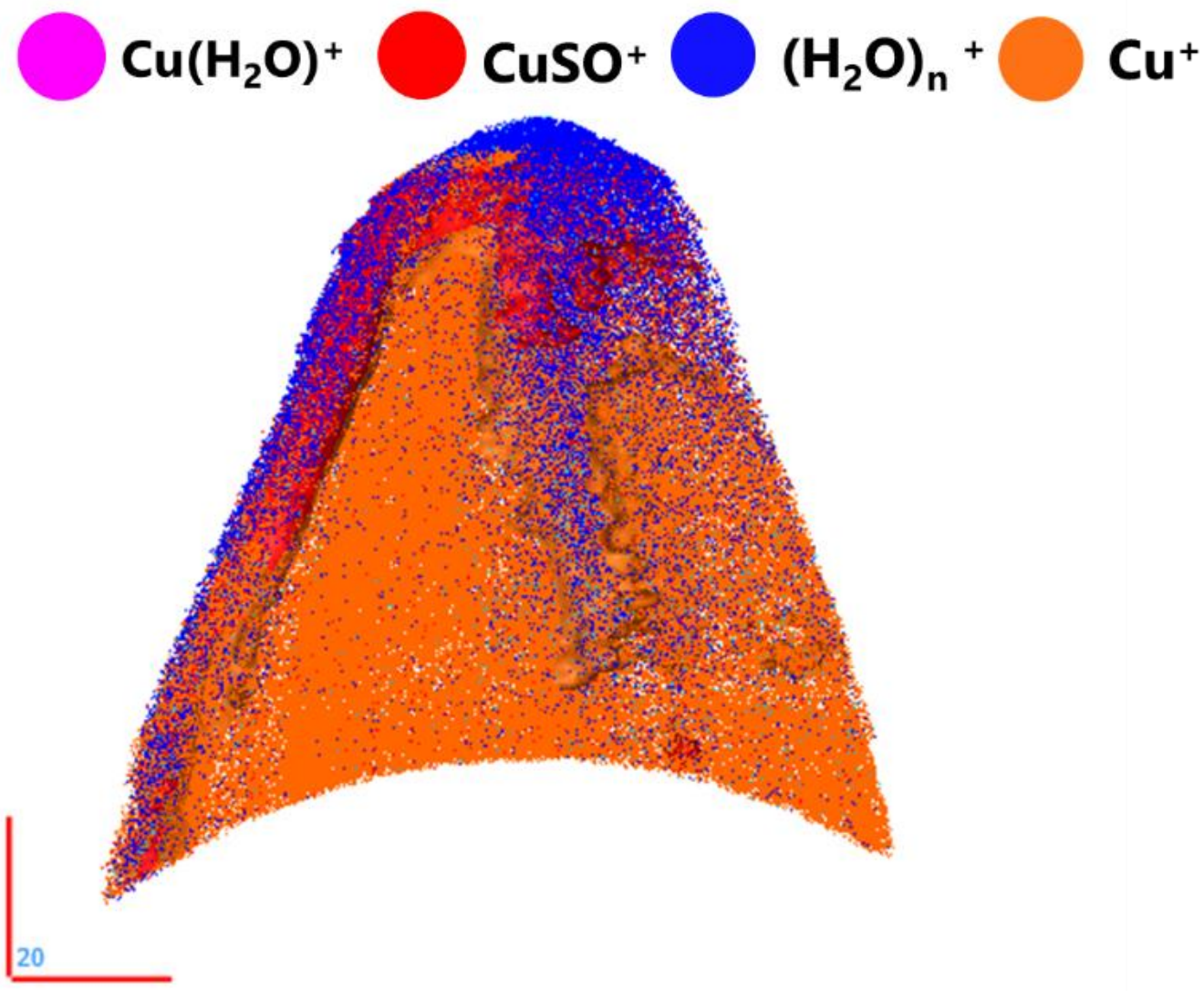

**Figure S6** Through-thickness cross-section parallel to the APT needle for the microcorrosion cell fabricated and cryo-FIB milled after exposure for 8 weeks

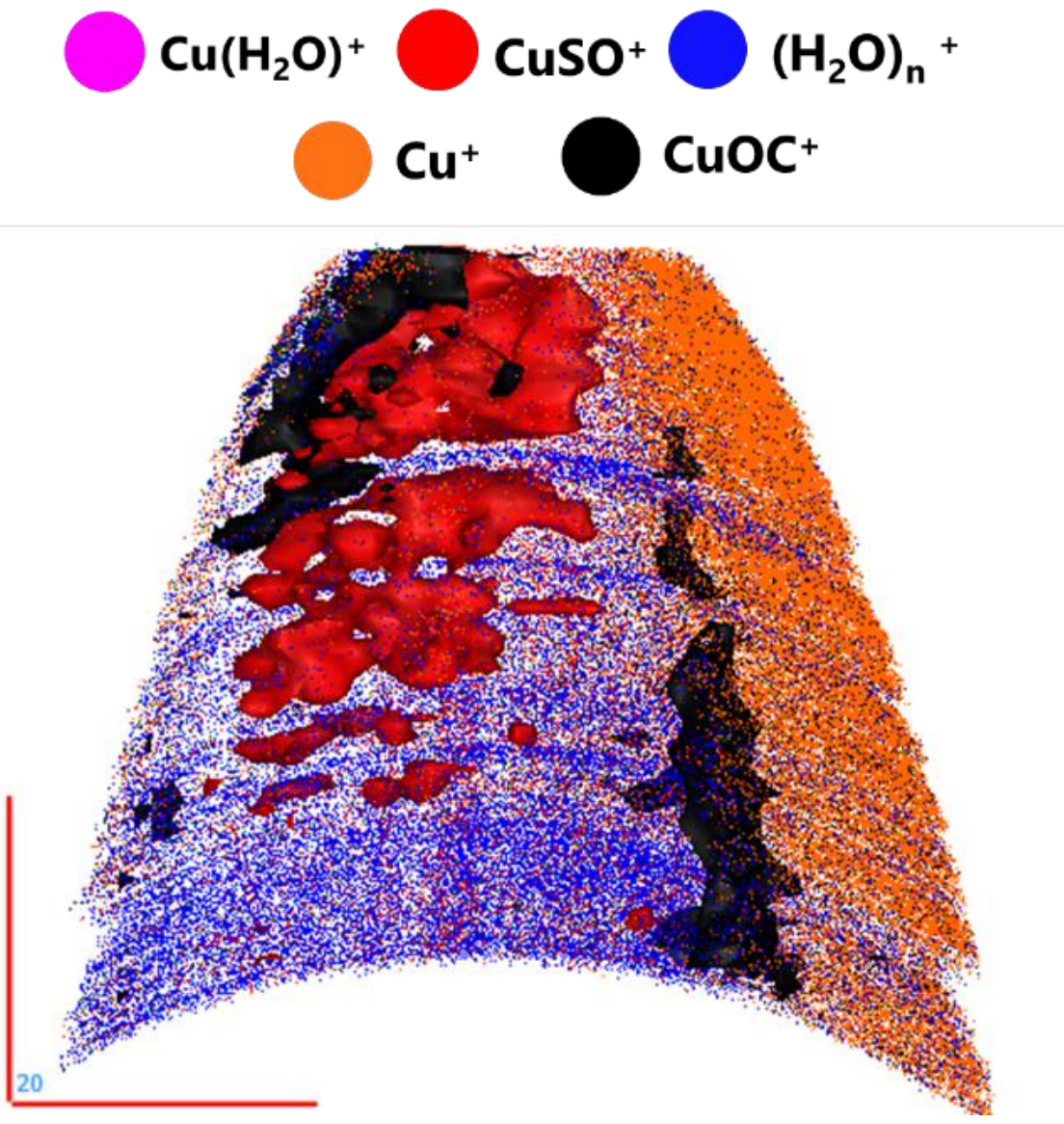

**Figure S7** Through-thickness cross-section parallel to the APT needle for the microcorrosion cell fabricated and cryo-FIB milled after exposure for 8 weeks and heated at 390K for 20 minutes

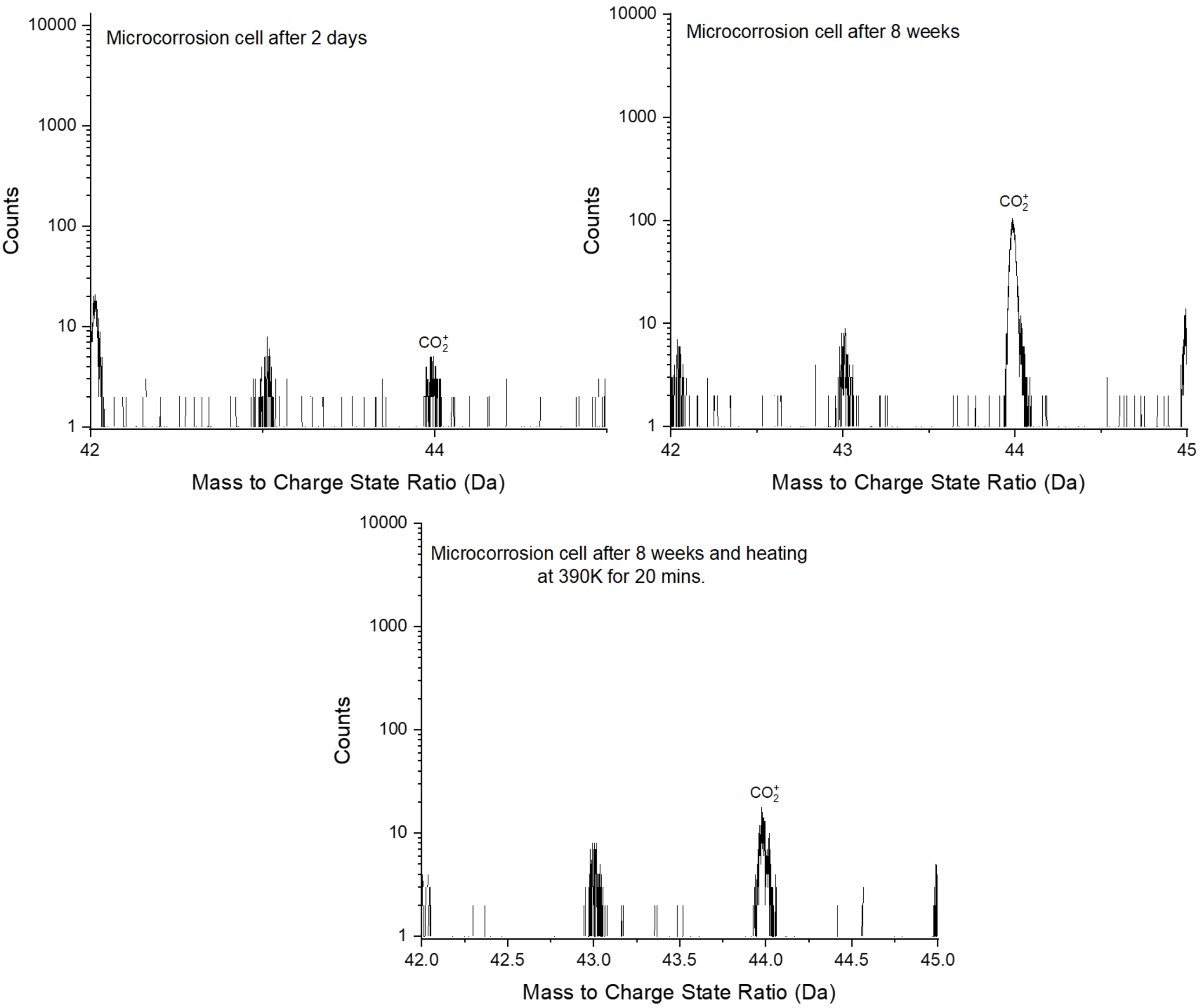

**Figure S8** The peak attributable to $CO_2^+$ is observable in the mass spectra for all the tested microcorrosion cell samples.

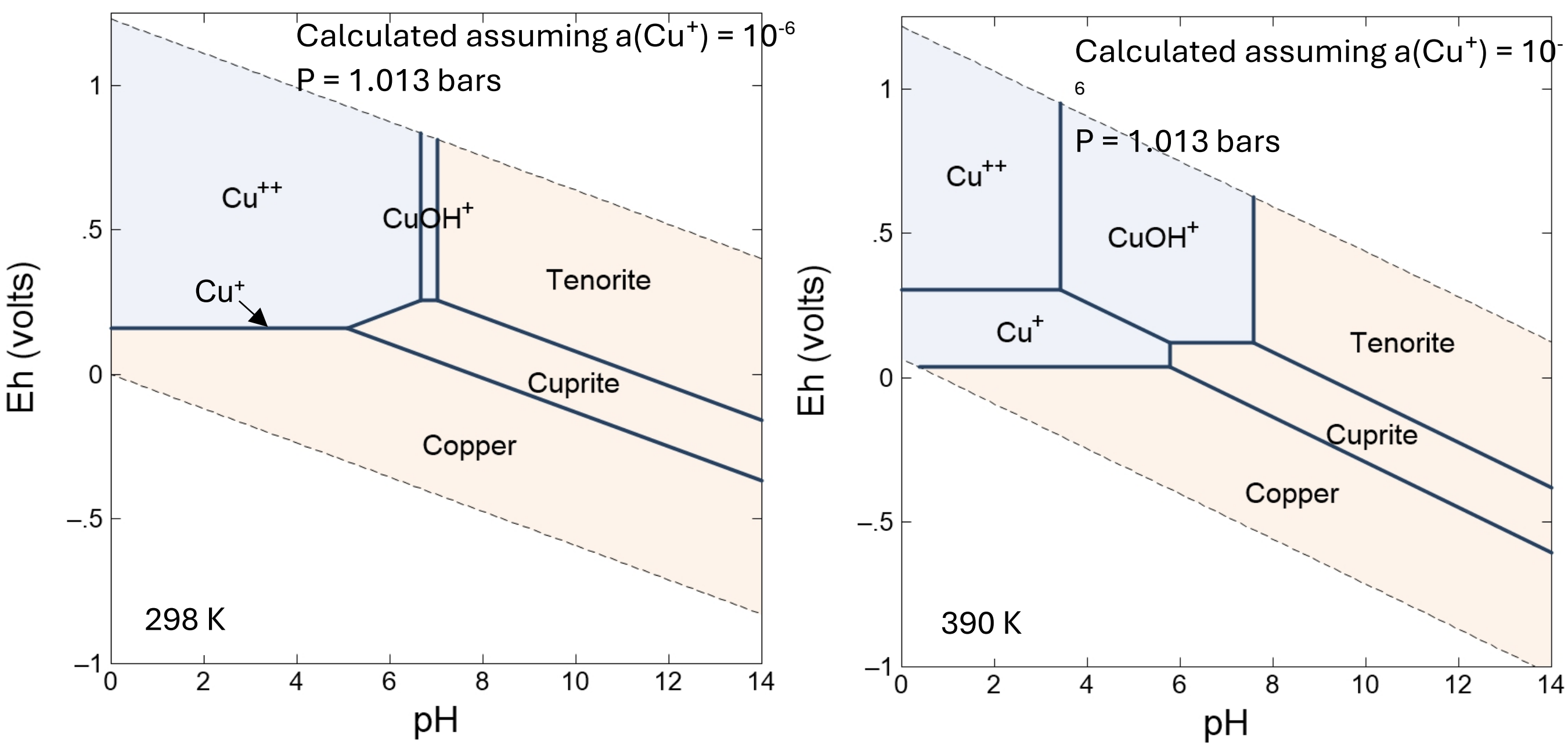

**Figure S9** Cu–$H_2O$ Pourbaix diagrams at 298 K and 390K.